\documentclass[aps,prc,showpacs,amsmath,amssymb]{revtex4}

\usepackage{epsfig}

\begin{document}

%\preprint{BNL-62600}

\title{ Cronin Effect and High $p_\perp$ Suppression in D+Au Collisions}

\author{D.~E.~Kahana$^{2}$, S.~H.~Kahana$^{1}$}
%\author{D.~E.~Kahana}
%\author{S.~H.~Kahana}

\address{$^{1}$Physics Department, Brookhaven National Laboratory\\
   Upton, NY 11973, USA\\
   $^{2}$31 Pembrook Dr., \\
   Stony Brook, NY 11790, USA}

\date{\today}  
 
\begin{abstract}
  
Great interest  has attached to  recent D+Au, $\sqrt{s} =  200$ A
GeV  data at  RHIC, obtained  with the  BRAHMS  detector. Between
pseudorapidities   $\eta=0$  and  $\eta=3.2$   the  appropriately
defined ratio  R[DAu$/$PP], comparing transverse  momentum spectra
of D+Au  to P+P,  exhibits a steady  decrease with  $\eta$.  This
diminuition is  examined within a two-stage  simulation, the last
stage  being  a purely  hadronic,  reduced  energy cascade.   The
result  is an  adequate  description of  the  data including  the
so-called  Cronin effect.  Additionally  there is  clear evidence
for  suppression,  in  the   second  stage,  of  relatively  high
transverse  momentum  $\eta=0$ leading  mesons,  {\it i.~e.}  the
Cronin effect,  only near mid  rapidity, is appreciably  muted by
final state interactions.

\end{abstract}


\maketitle 
\section{Introduction}

It is instructive to begin  with a comparison of the minimum-bias
experimental  distribution $dN^{ch}/d\eta$ for  charged particles
at  $\sqrt{s}=200$ A  GeV and  our eventual  attempt  to describe
this.  These are shown in  Fig(1) together with the equivalent PP
measurement  without any normalising  factors imposed.   The data
shown are from  UA5~\cite{ua5} and PHOBOS~\cite{phobos1}. Clearly
charged  particle  production in  D+Au  is considerably  enhanced
relative  to   PP;  but  for  large  $\eta$   the  D+Au  spectrum
asymptotically joins PP, both in  the data and in our simulation.
It is also evident the $\eta=3.2$ point is appreciably suppressed
relative to  $\eta=0$, for D+Au but  not for PP.   It will become
clear  that  these  pseudorapidity  spectra, which  are  in  fact
integrals  of  the  double differential  cross-sections  $(1/2\pi
p_\perp)[d^2N^{ch}/dp_\perp\,d\eta]$,  are   dominated  by  quite
small transverse momenta $p_\perp$.  We examine this self-evident
thesis   in   more   detail    as   we   proceed.    

The BRAHMS collaboration~\cite{brahms1} has focused on the $\eta$
and $p_\perp$-dependence of the ratio

\begin{equation}
\text{R}[\text{DAu/PP}] =
\left(\frac{1}{\text{N}_{coll}}\right)
\frac{
[d^2N^{ch}/dp_\perp\,d\eta]\,\,({\text{DAu}})
}
{
[d^2N^{ch}/dp_\perp\,d\eta]\,\,({\text{PP}})
}
,
\end{equation}

\noindent  where  $\text{N}_{coll}$ is  a  calculated number  of
binary  NN  collisions   occurring  in  minimum-bias  D+Au.  They
consider also  the combined  $\eta$ and $p_\perp$  dependence for
varying centralities $c_{1,2}$ of

\begin{equation}
\text{R}_{c_ip}= \left[ \frac{\text{R}_{c_i}[{\text{DAu/PP}}]}{\text{R}_{p}[\text{{DAu/PP}}]} \right],
\end{equation}
where the denominator is the same ratio for a peripheral setting.
The behaviour of the former ratio we contend is determined mainly
by  the   low  transverse  momentum  dynamics   and  by  $p_\perp$
distribution in  PP collisions, which is  for us an  input to the
nucleus-nucleus   simulation.    There   are  important   dynamic
modifications,  e.~g. the  Cronin  effect~\cite{cronin}, but  the
relation  between low  and high  $p_\perp$ is  to a  large extent
similar   to   that   in   PP.    The   observed   behaviour   of
$\text{R}_{c_ip}$   with   pseudorapidity   and   centrality   is
determined by  the asymptotic approach  of $dN/d\eta$ in  D+Au to
that  in PP at  increasing $\eta$,  and the  resulting diminished
variation with $\eta$.

The code LUCIFER, developed  for high energy heavy-ion collisions
has    previously   been   applied    to   both    SPS   energies
$\sqrt{s}=(17.2,20)$ A  GeV~\cite{lucifer1} and to  RHIC energies
$\sqrt{s}=(56,130,200)$ A  GeV ~\cite{lucifer2,luc3}.  We present
a  brief  description  of   the  dynamics  of  this  Monte  Carlo
simulation.  Many other simulations of heavy ion collisions exist
and  these are  frequently  hybrid in  nature,  using say  string
models                in                the               initial
state~\cite{rqmd,rqmd2,frithjof,bass1,capella,werner,ko,ranft}
together with  final state hadronic collisions,  while some codes
are   purely   partonic~\cite{boal,eskola,wang,geiger,bass2}   in
nature.  Our  approach is closest in  spirit to that  of RQMD and
K.~Gallmeister,  C.~Greiner,  and  Z.~Xu  and  parallel  work  by
W.~Cassing~\cite{rqmd,greiner,cassing}.   Certainly  our  results
seem to  parallel those  of the latter  authors.  

The  purpose of  describing such  high energy  collisions without
introducing the  evidently existing parton nature  of hadrons, at
least  for soft  processes, was  to  set a  baseline for  judging
whether  deviations from the  simulation measured  in experiments
existed  and  could   then  signal  interesting  phenomena.   The
division  between soft and  hard processes,  the latter  being in
principle described by perturbative  QCD, is not necessarily easy
to identify  in heavy ion  data.  For the relatively  simple D+Au
system we are interested in  separating the effects of our second
stage, a lower  energy hadronic cascade, from those  of our first
stage,  a parallel  rather than  sequential treatment  of initial
(target)-(projectile) NN interactions.

\section{The Simulation}
\subsection{Stage I}

The first  stage I of LUCIFER considers  the initial interactions
between the separate  nucleons in the colliding ions  A+B, but is
not a cascade.  The  totality of events involving each projectile
particle  happen  essentially  together   or  one  might  say  in
parallel.   Neither  energy   loss  nor  creation  of  transverse
momentum  ($p_\perp$)  are  permitted  in  stage  I,  clearly  an
approximation.           A          model          of          NN
collisions~\cite{lucifer1,lucifer2},   incorporating  most  known
inclusive cross-section and multiplicity data, guides stage I and
sets up the initial conditions for stage II.  The two body model,
clearly an  input to  our simulation, is  fitted to  the elastic,
single   diffractive  (SD)   and  non-single   diffractive  (NSD)
aspects~\cite{goulianos}   of   high   energy   $PP$   collisions
~\cite{ua5,ua1}  and  $P\bar   P$  data~\cite{fermilab}.   It  is
precisely  the  energy   dependence  of  the  cross-sections  and
multiplicities  of  the  NN  input  that led  to  our  successful
prediction~\cite{lucifer2,phobos1}  of the rather  small ($13\%$)
increase   in    $dN^{ch}/d\eta$   between   $\sqrt{s}=130$   and
$\sqrt{s}=200$ A GeV, seen in the PHOBOS data~\cite{phobos3}.

A history of  the collisions that occur between  nucleons as they
move along straight  lines in stage I is  recorded and later used
to guide determination  of multiplicity.  Collision driven random
walk  in $p_\perp$  fixes the  $p_\perp$  to be  ascribed to  the
baryons  at the  start of  stage II.   The  overall multiplicity,
however, is  subject to a  modification, based, as we  believe on
natural physical requirements~\cite{lucifer2}.

If a sufficiently hard  process, for example Drell-Yan production
of  a lepton  pair at  large mass  occurred, it  would lead  to a
prompt  energy loss  in stage  I.  Hard  quarks and  gluons could
similarly be  entered into the particle lists  and their parallel
progression followed.  This has not yet been done.  One viewpoint
and justification for our approach is to say we attempt to ignore
the direct effect  of colour on the dynamics,  projecting out all
states  of the  combined  system possessing  colour.   In such  a
situation  there  should  be  a duality  between  quark-gluon  or
hadronic treatments.
 
The  collective/parallel method  of treating  many  NN collisions
between the target and projectile is achieved by defining a group
structure for  interacting baryons.  This is  best illustrated by
considering a prototype  proton-nucleus (P+A) collision.  A group
is  defined  by spatial  contiguity.   A  proton  at some  impact
parameter  $b(\bar{x}_\perp)$  is  imagined  to  collide  with  a
corresponding  `row'  of   nucleons  sufficiently  close  in  the
transverse  direction to the  straight line  path of  the proton,
{\it   i.~e.}~within   a  distance   corresponding   to  the   NN
cross-section.   In   a  nucleus-nucleus  (A+B)   collision  this
procedure is generalized by making  two passes: on the first pass
one includes all  nucleons from the target which  come within the
given  transverse distance  of some  initial  projectile nucleon,
then on the  second pass one includes for  each target nucleon so
chosen, all of those  nucleons from the projectile approaching it
within the  same transverse distance.  This  totality of mutually
colliding   nucleons,   at   more   or   less   equal   transverse
displacements,  constitute  a  group.  The  procedure  partitions
target and projectile nucleons into a set of disjoint interacting
groups as well as a set of non-interacting spectators in a manner
depending on the overall  geometry of the A+B collision.  Clearly
the largest groups in P+A will,  in this way, be formed for small
impact parameters  $b$; while for the  most peripheral collisions
the groups  will almost always  consist of only one  colliding NN
pair. Similar conclusions hold in the case of A+B collisions.
     
In stage II of the  cascade we treat the entities which rescatter
as  prehadrons.  These  prehadrons, both  baryonic or  mesonic in
type, are not the  physical hadron resonances or stable particles
appearing in  the particle  data tables, which  materialise after
hadronisation.   Importantly prehadrons  are allowed  to interact
starting   at    early   times,   after    a   short   production
time~\cite{boris}, nominally  the target-projectile crossing time
$T_{AB} \sim R_{AB}/\gamma$.  The mesonic prehadrons are imagined
to have  ($q \bar  q$) quark content  and their  interactions are
akin to  the dipole interactions included in  models relying more
closely  on explicit  QCD~\cite{boris,mueller},  but are  treated
here  as colourless  objects.   

Some  theoretical  evidence   for  the  existence  of  comparable
colourless structures is  given by Shuryak and Zahed~\cite{zahed}
and  by certain lattice  gauge studies~\cite{lattice}.   In these
latter  works  a basis  is  established  for  the persistence  of
loosely  bound  or  resonant   hadrons  above  the  QCD  critical
temperature $T_c$ to $T  \sim (1.5-2.0)\times T_c$.  This implies
a persistence to much higher transverse energy densities $\rho(E)
\sim (1.5-2.0)^4  \rho_c$, hence  to the early  stages of  a RHIC
collision.  Accordingly  we have incorporated into  stage II {\it
hadron  sized}  cross-sections  for  the  interactions  of  these
prehadrons,  altough early  on it  may  in fact  be difficult  to
distinguish  their colour  content.   Such larger  cross-sections
indeed  appear  to  be  necessary  for  the  explanation  of  the
apparently   large    elliptical   flow   parameter    found   in
measurements~\cite {molnar,flow}.

The  prehadrons when mesonic  may consist  of a  spatially close,
loosely correlated  quark and anti-quark  pair, are given  a mass
spectrum between $m_\pi$ and $1$ GeV, with correspondingly higher
upper and  lower limits allowed for  prehadrons including strange
quarks.  The Monte-Carlo selection  of masses is then governed by
a Gaussian distribution,
\begin{equation}
P(m)= exp(-(m-m_0)^2/w^2),
\end{equation}
with $m_0$ a selected  center for the prehadron mass distribution
and $w=m_0/4$  the width.  The non-strange  mesonic prehadrons is
taken at  $m_0 \sim  500$ MeV, and  for strange at  $m_0\sim 650$
MeV. Small changes in $m_0$  and $w$ have little effect since the
code is constrained to fit hadron-hadron data.

Too high an  upper limit for $m_0$ would  destroy the soft nature
expected for most prehadron interactions when they finally decay
into  `stable'  mesons.   The   same  proviso  is  in  place  for
prebaryons which are restricted to a mass spectrum from $m_N$ to
$2$ GeV.   However, in  the present calculations  the prebaryons
are  for simplicity  taken just  to be  the normal  baryons.  The
mesonic  prehadrons  have   isospin  structure  corresponding  to
$\rho$, $\omega$,  or $K^*$, while  the baryons range  across the
octet and decuplet.

Creating  these intermediate  degrees of  freedom at  the  end of
stage I  simply allows the original nucleons  to distribute their
initial energy-momentum  across a larger basis of  states or Fock
space, just  as is done in  string models, or for  that matter in
partonic   cascade   models.    Eventually,  of   course,   these
intermediate  objects decay  into physical  hadrons and  for that
purpose we assign a  uniform decay width $\sim \Gamma_{f}$, which
then plays the role of a hadronisation or formation time.

\subsection{Groups}
Energy  loss  and  multiplicity  in  each group  of  nucleons  is
estimated from  the straight line collision  history.  To repeat,
transverse momentum  of prebaryons is  assigned by a  random walk
having  a number  of  steps  equal to  the  number of  collisions
suffered.   The  multiplicity  of  mesonic prehadrons  cannot  be
similarly directly estimated from  the number of NN collisions in
a group.   We argue~\cite{gottfried} that  only spatial densities
of generic prehadrons~\cite{lucifer1,lucifer2} below some maximum
are  allowable,  {\it  viz.}   the prehadrons  must  not  overlap
spatially at the  beginning of stage II of  the cascade.  The KNO
scaled multiplicity distributions, present  in our NN modeling are
sufficiently  long-tailed  that imposing  such  a restriction  on
overall multiplicity can for larger nuclei affect results even in
P+A  or D+A  systems.  In  earlier  work~\cite{lucifer1,luc3} the
centrality dependence of $dN/d\eta$ distributions for RHIC energy
Au+Au  collisions   was  well  described  with   such  a  density
limitation on the prehadrons.  

Importantly,  the cross-sections  in prehadronic  collisions were
assumed to be the same size as hadronic, {\it e.~g.}~meson-baryon
or meson-meson  {\it etc.}~, at  the same center of  mass energy,
thus introducing  no additional  free parameters into  the model.
Where the  latter cross-sections or their  energy dependences are
inadequately known we  employed straightforward quark counting to
estimate the  scale. In both SPS  Pb+Pb and RHIC  Au+Au events at
several energies it was sufficient to impose this constraint at a
single energy.  The inherent energy dependence  in the KNO-scaled
multiplicities of the NN inputs and the geometry then take over.

\subsection{High Transverse Momenta}
One  question which  has yet  to be  addressed concerns  the high
$p_\perp$  tails  included in  our  calculations.  In  principle,
LUCIFER  is  applicable  to  soft processes  {\it  i.~e.}~at  low
transverse momentum.  Where the cutoff in $p_\perp$ occurs is not
readily apparent. In any case we can include high $p_\perp$ meson
events through  inclusion in the  basic hadron-hadron interaction
which  is  of  course  an  input  rather than  a  result  of  our
simulation.   Thus   in  Fig(2)  we  display   the  NSD  $(1/2\pi
p_\perp)(d^2N^{charged}/dp_\perp\,d\eta)$   from  UA1~\cite{ua1}.
One can use  a single exponential together with  a power-law tail
in $p_\perp$, or alternatively two exponentials, to achieve a fit
of the  output in PP to  UA1 $\sqrt{s}$=200 GeV  data. A sampling
function of the form
\begin{equation}
f = p_\perp (a exp(-p_\perp/w) + b / ((1 + (r / \alpha)^ \beta)),    
\end{equation}
gives a  satisfactory fit to the PP data in the Monte-Carlo.
  
This  PP $p_\perp$  spectrum,  inserted into  the  code, is  then
applied  to  the  meson   $p_\perp$  distribution  in  D+Au.   No
correction  is made  for  possible  energy loss  in  stage I,  an
assumption parallel to that made by the BRAHMS and all other RHIC
experiments,  in analysing  $p_\perp$ spectra  and multiplicities
irrespective   of  low   or   high  values.    Since  we   impose
energy-momentum  conservation  in each  group,  a high  $p_\perp$
particle having  say, several GeV/c of  transverse momentum, must
be  accompanied in the  opposite transverse  direction by  one or
several compensating  mesons.  Such high-$p_\perp$  particles are
not exactly  jets, to the extent  that they did  not originate in
our simulation from hard parton-parton collisions, but they yield
the same observable experimental behaviour.

\subsection{Initial Conditions for II}
The final operation  in stage I is to  set the initial conditions
for the hadronic cascade  in stage II.  The energy-momentum taken
from the initial baryons and shared among the produced prehadrons
is  established  and an  upper  limit  placed  on the  production
multiplicity   of  prehadrons  and   normal  hadrons.    A  final
accounting of  energy sharing is  carried out through  an overall
4-momentum conservation  requirement.  We emphasize  that this is
carried out separately within each group of interacting nucleons.

The spatial  positioning of the  particles at this time  could be
accomplished in a  variety of ways.  We have  chosen to place the
prehadrons in each group  inside a cylinder, initially having the
longitudinal size of the nucleus, for a P+A collision, and having
the longitudinal size of  the interaction region at time $T_{AB}$
in an A+B collision, then  allowing the cylinder to evolve freely
according to the longitudinal momentum distributions, for a fixed
time $\tau_{f}$, defined in the rest frame of each group.  At the
end of $\tau_f$ the multiplicity  of the prehadrons is limited so
that,  if given  normal  hadronic sizes  $\sim (4\pi/3)  (0.8)^3$
fm$^{3}$, they do not overlap within the cylinder.

Up  to this  point  longitudinal boost  invariance is  completely
preserved,  since stage  I  is carried  out  using straight  line
paths.  The technique of defining the evolution time in the group
rest frame  is essential to minimizing  residual frame dependence
which inevitably arises in any cascade, hadronic or partonic,when
transverse momentum is  considered due to the finite  size of the
colliding   objects  implied   by   their  non-zero   interaction
cross-sections.
       
\section{Stage II: Final State Cascade}
Stage  II is  as stated  a straightforward  cascade in  which the
prehadronic  resonances  interact  and  decay as  do  any  normal
hadrons  present or  produced during  this  cascade.  Appreciable
energy having being finally transferred to the produced particles
these  `final  state' interactions  occur  at considerably  lower
energy than the initial nucleon-nucleon collisions of stage I. As
pointed out,  during stage II  the interaction and decay  of both
prehadrons  and  hadrons  is  allowed.   In the  case  of  D+Au,
although less abundant than with a more massive projectile, these
final state  interactions are as we will  see, nevertheless still
of some relevance.

We are then in a position to present results for D+Au collisions.
These  appear   in  Fig(1),   as  previously  referred   to,  and
Figs(2--6), some  of which are comparisons  with the measurements
of both  BRAHMS and PHOBOS~\cite{brahms1, phobos1}.   In fact the
plot    of    experimental     data    in    Fig(1)    is    from
PHOBOS~\cite{phobos1}.   This   PHOBOS  reference  also  exhibits
comparisons           with           several          theoretical
calculations~\cite{rqmd2,cgc1,wang2,ampt}.     Two    of    these
references~\cite{wang2,ampt},  describe   much  of  the  measured
$\eta$-distribution  at negative  $\eta$ near  the  target, while
one~\cite{ampt}  apparently  accounts  for the  extreme  backward
tail; this is a subject to which we will return.

The initial  conditions created to start the  final cascade could
have perhaps  been arrived  at through more  traditional, perhaps
partonic, means.  The second stage would then still proceed as it
does here.  We reiterate that  our purpose has been to understand
to what extent the results  seen in Figures (1-7) are affected by
stages  I and  II  separately.  {\it  i.~e.}~do  they arise  from
initial or from final state interactions.

\section{Results: Comparison with Data}
Fig(3) contains the  simulated charged transverse momenta spectra
for  D+Au at  $\eta=(0,3.2)$ alongside  the UA1  data.   The many
orders of  magnitude fall in $p_\perp$  densities with rapidity
is  apparent.   Aside from  low  $p_\perp$  the  D+Au curves  for
increasing  $p_\perp$ appear  roughly parallel  to PP;  small but
interesting deviations show up when the ratios previously defined
are displayed.   Additionally, direct integration  of the spectra
indicates  that $  \ge 90\%$  of the  charged  rapidity densities
result from $p_\perp  \le 0.7$ GeV/c.  Having built  in no energy
loss  effects on  these  $p_\perp$ distributions  in the  initial
state, that a similar fall off obtains in both D+Au and PP is all
but  preordained, and  the overall  ratio between  $\eta=0.0$ and
$\eta=3.2$  seems  to  be  driven  completely  by  low  $p_\perp$
physics.  In fact the multiplicity choice at given pseudorapidity
and transverse  momentum is  only mildly influenced  by $p_\perp$
dependence aside from that already present in the PP input.

In Fig(4) the calculated LUCIFER ratios $\text{R}[\text{DAu/PP}]$
are  plotted alongside  those for  BRAHMS~\cite{brahms1}  at both
$\eta=0$  and $\eta=3.2$.  The  theoretical results  are obtained
using  $N_{coll}=7.0$  rather  than  the value  closer  to  $7.2$
employed  by BRAHMS.  Our  calculation of  the average  number of
collisions in  minimum-bias D+Au,  defined as $b  \le 16$  fm, is
approximately  $7.0$.   The  Cronin  affect  is  evident  in  the
calculated $\eta=0$  spectrum, less  so for $\eta=3.2$.   This is
not unexpected.

\subsection{Jet Suppression}
A very  interesting result is  obtained by turning off  the final
cascade, {\it  i.~e.}~stage II.  Then the  prehadrons produced in
stage  I evolve  or decay  into stable  particles after  the time
$\tau_f \sim 1/(\Gamma)$ and do not otherwise suffer interaction.
This situation is described in Fig(5), where it is clear that the
magnitude   of  the  Cronin   enhancement  of   $dN/dp_\perp$  is
considerably  magnified. This  enhancement  is then  very much  a
creature of I, i.~e. a  product of the transverse momentum gained
in collisions  with nucleons in the  target. Incidentally, Fig(5)
also  indicates  that   a  compensating  increase  in  transverse
momentum density occurs at the lowest $p_\perp$.

One might well turn this  around and declare that the final state
scattering of  a given prehadron  with comovers has cut  down the
Cronin effect.   This is an appreciable  reduction which suggests
the  applicability of  the  term `jet  suppression', a  reduction
which indeed constitutes final  state suppression.  The change in
$p_\perp$ spectra  is considerably less  for $\eta \sim  3$ where
Fig(1)  indicates considerately  less comovers  are  present, and
indeed the Cronin enhancement is less evident at the more forward
$\eta$.

The spillover of such stage II comovers decreases with increasing
distance from  the target  pseudorapidities.  The Cronin  rise is
itself    less   evident   at    forward   $\eta$,    the   ratio
$\text{R}[\text{DAu/PP}]$ is  flatter as a  function of $p_\perp$.
Indeed as $\eta$ increases  the $p_\perp$ spectra approach to the
PP  spectra.   This  is  again easily  understandable:  the  most
peripheral collisions involving  the least number of participants
will contribute  more strongly to more  forward rapidities.  Both
the unrenormalised theoretical and measured D+Au curves appear to
merge with  PP at the  largest pseudorapidities shown  in Fig(1).
One  expects to  see  a corresponding  behaviour with  decreasing
centrality and decreasing participant nucleon number.

\subsection{Centrality Dependence of R[DAu/PP]}
A similar  theme then is  repeated in Fig(6) where  the $p_\perp$
spectra  for two  quite  different degrees  of centrality  differ
markedly; the disparate centralities are  defined by $b \le 4$ fm
and $b=8$  fm.  The $\eta$  dependence exhibited for $b=8$  fm, a
clearly peripheral geometry, is very muted with both $\eta=0$ and
$\eta=3$ showing  a strong resemblance  to the PP  $\eta=0$ data.
The more  central choice, $b \le  4$, is subject  to quite strong
pseudo-rapidity variation.  This explains at least qualitatively,
the  behaviour  of  the  second  ratio BRAHMS  focuses  on,  {\it
i.~e.}~$\text{R}_{cp}$.~\cite{brahms1}.   The  crossover of  this
ratio with centrality as a function of $\eta$ is explained by the
much flatter dependence in  $\eta$ discussed here.  The differing
number  $\text{N}_{coll}$ then  must be  invoked to  complete the
picture but it  plays only a passive role,  the flattening of the
$p_\perp$-distribution  already  evident  in  Fig(6)  for  widely
differing  centralities or impact  parameter plays  the essential
role.

\section{Conclusions}

It is hard  to conclude definitively from what  is presented here
that  the gluon  saturation~\cite{saturation,cgc1}  and attendant
colour-glass-condensate  interpretation~\cite{cgc1,cgc2}  of  the
BRAHMS  data  is  not  a  more  fundamental  explanation  of  the
measurements  discussed here.   Certainly the  low-$x$  basis for
this  modeling  is  related  to  the  increasing  $\eta$  picture
presented here,  and perhaps the gluon saturation  aspect of that
approach   is   mirrored  in,   and   underpins,  the   prehadron
multiplicity limitation employed above.

It  would  seem  however  that  the  direct  attempt  at  a  PQCD
explanation of this behaviour must claim that, at the very least,
all soft mesons are produced in essentially hard collisions.  The
presentation  here provides  an interesting  case for  relying on
essentially soft,  low $p_\perp$, processes to  produce the major
features  of the BRAHMS  data.  True  enough, the  high $p_\perp$
tails in distributions are merely  tacked on in our approach, but
legitimately  so   by  using  the   PP  data  as  input   to  the
nucleus-nucleus  cascade.  The  PP  $dN/dp_\perp$ variation  with
$p_\perp$ to a large extent drives the multiplicity generation in
D+Au, altering only slightly the hard $\eta$-dependent ratio from
the soft.  One only  need add the  assumption that  the $p_\perp$
tails in D+Au do not exhibit any drastic non-monotonic behaviour.
 
The nascent appearance of appreciable high $p_\perp$ suppression,
especially in our $\eta$=0  spectrum seen in Fig.5, suggests that
enhanced    suppression   will    occur   in    a    full   Au+Au
collision. Whether  the complete, or an  appreciable fraction of,
jet  suppression seen  in RHIC  experiments can  be  explained by
final state interactions remains  to be established. We note that
C.~Greiner         and        coauthors~\cite{greiner}        and
Cassing~\cite{cassing}  have  commented  forcefully on  precisely
this point, also in a hadron-based cascade setting.

We indicated we would return to the target pseudorapidity region.
It is of direct relevance  to do this for discussion of specifics
in D+Au but also for the implications for Au+Au and other complex
systems. For the deuteron projectile PHOBOS data not only extends
further backward than other experiments or calculations, but also
exhibits  a   feature,  perhaps   a  shoulder,  near   where  our
calculations exhibit a peak  in the charged baryons (see Fig(1)).
It   would  clearly   be  of   some  import   to   have  particle
identification  in   the  present  measurements   of  $\eta$  and
$p_\perp$ distributions.  Considering  the D+Au system, one notes
that  transverse  momentum  distributions  near  the  target,  or
further back in $\eta$,  are significantly softer, again possibly
anticipating high $p_\perp$ suppression to be associated with the
symmetric massive ion collisions.

\vfill\eject

\begin{figure}
\vbox{\hbox to\hsize{\hfil
\epsfxsize=6.1truein\epsffile[0 0 561 751]{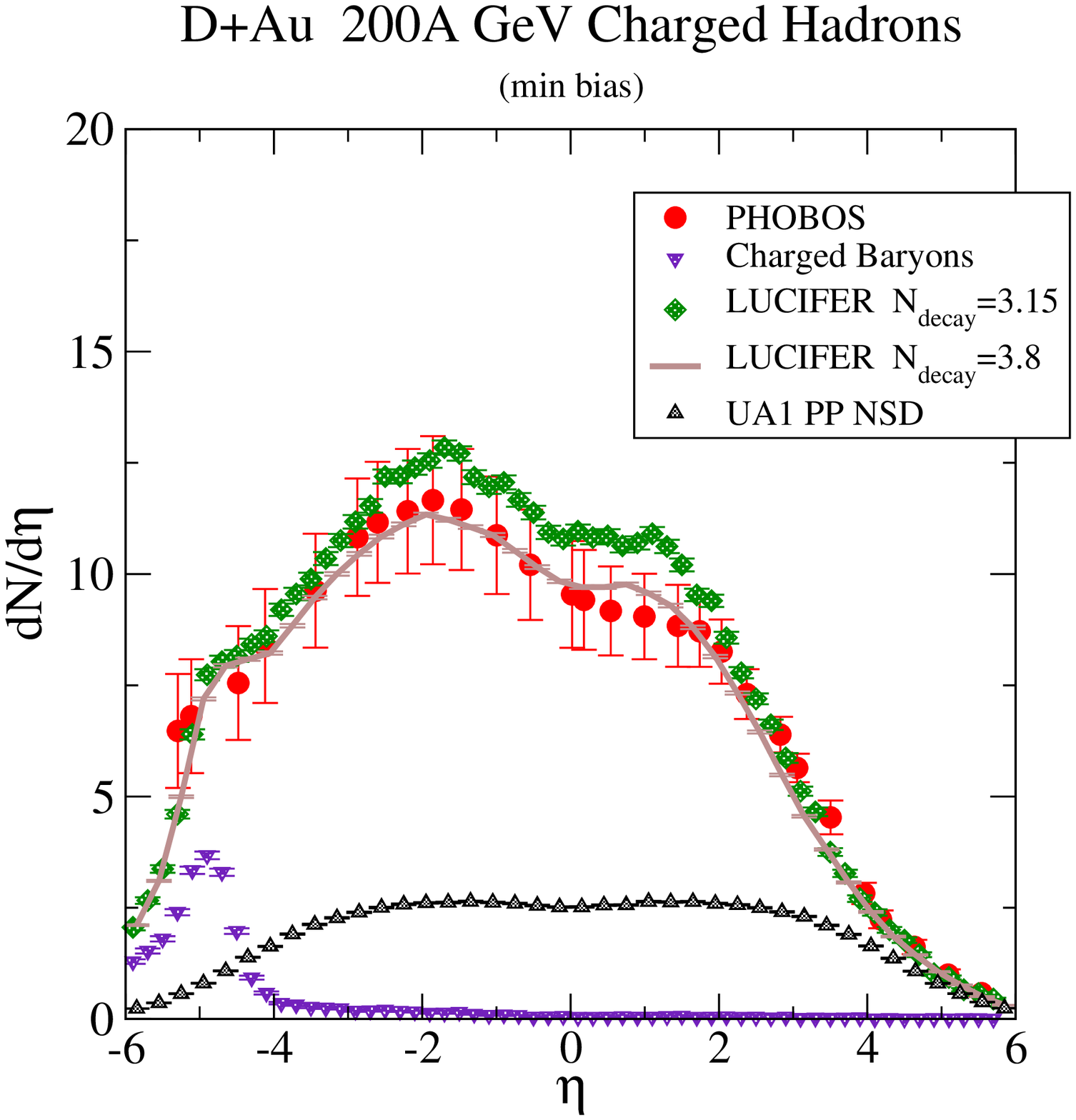}
\hfil}}
\caption[]{D+Au  Pseudorapidity  spectra:  Direct  comparison  of
PHOBOS minimum bias data  with LUCIFER simulation, the latter for
$b le  16$ fm.  Two calculations are  shown, slightly differently
normalized,  one  for which  the  average  prehadron decays  into
$3.15$ observed  mesons and one  for which this number  is $3.8$.
Both  are   only  slightly  above  that  expected   for  pure  NN
production,   indicating  some   calculated  events   exceed  the
multiplicity constraint  discussed in  the text.  The  absence of
collision  number  divisors is  instructive,  revealing both  the
considerable  production of  final state  mesons at  $\eta$=0, in
excess of PP,  and the apparent asymptotic approach  of both data
and calculation to PP at the largest observed $\eta$.}
\label{fig:Fig.1}
\end{figure}
\clearpage

\begin{figure}
\vbox{\hbox to\hsize{\hfil
\epsfxsize=6.1truein\epsffile[0 0 561 751]{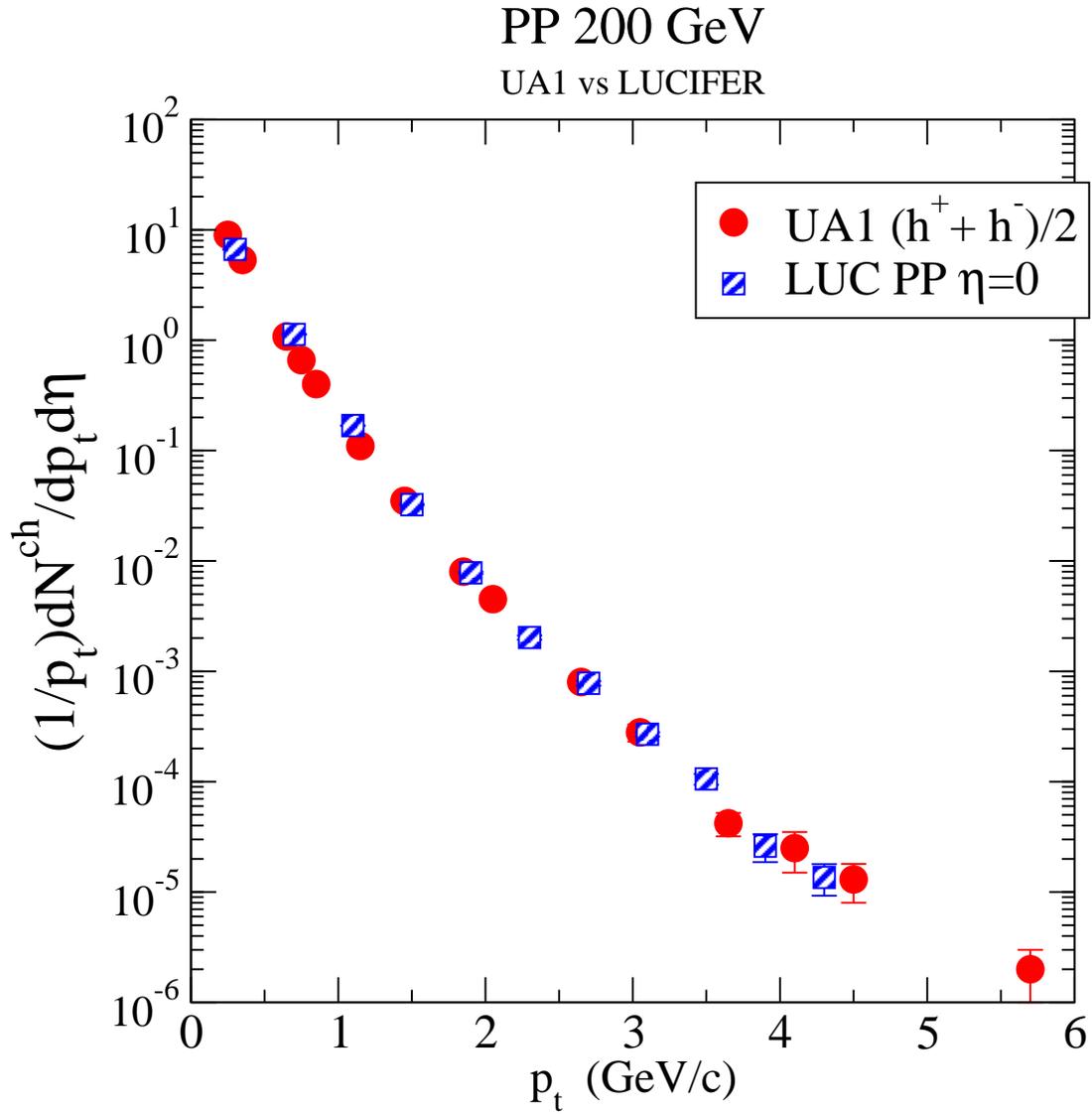}
\hfil}}
\caption[]{Transverse Momentum Spectra:  UA1 vs LUCIFER. The basic
input of both low and  high $p_t$ for simulations obtained from a
fit to the SPS data from the UA1 Collaboration~\cite{ua1}.}
\label{fig:Fig.2}
\end{figure}
\clearpage

\begin{figure}
\vbox{\hbox to\hsize{\hfil
\epsfxsize=6.1truein\epsffile[0 0 561 751]{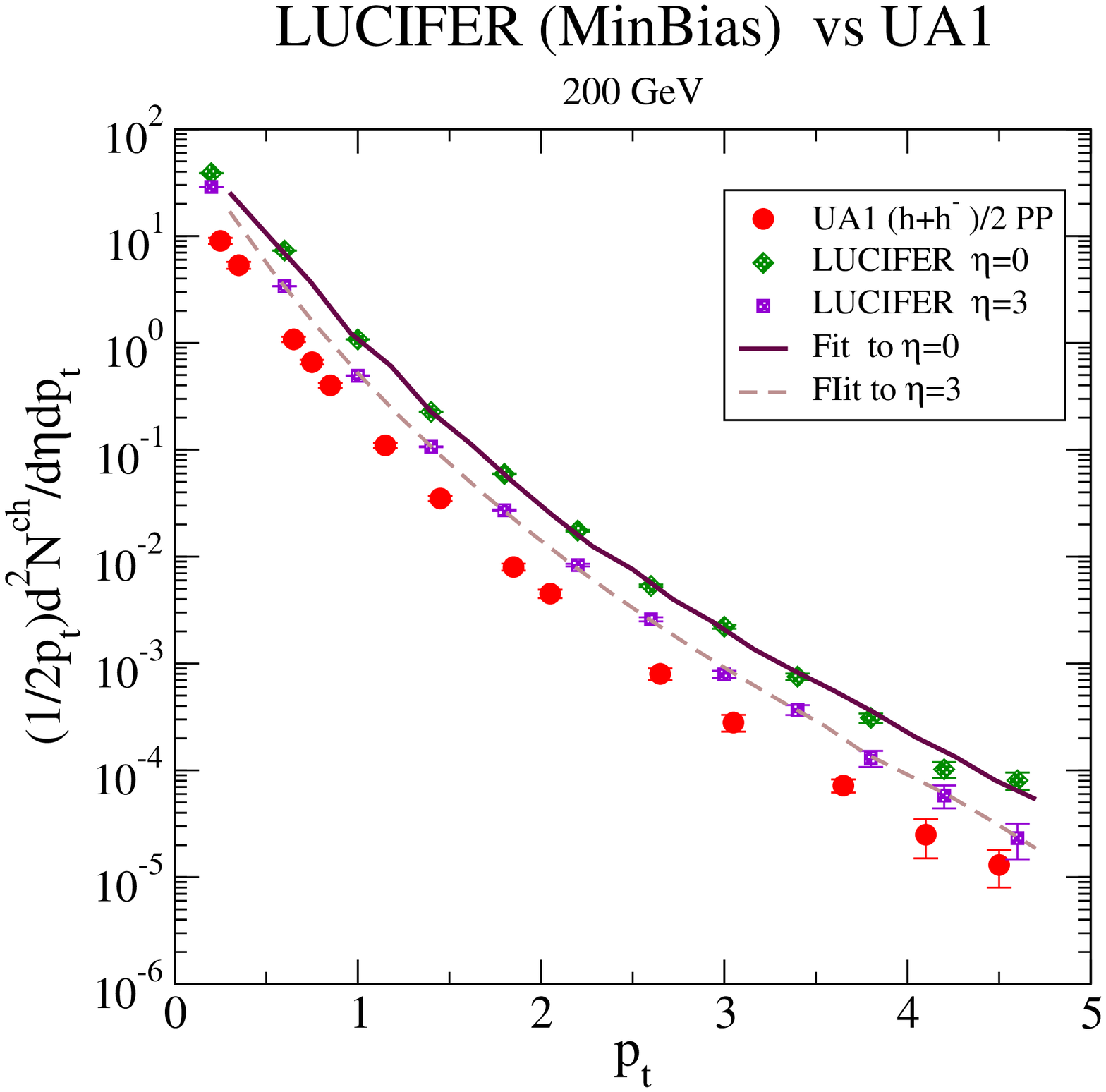}
\hfil}}
\caption[]{The simulated  charged transverse momenta  spectra for
$D+Au$  at $\eta=(0,3.0)$  alongside the  UA1 data.  Fits  to the
LUCIFER  calculations,  used to  interpolate  the simulation  are
shown.  These   are  made  with  the  combination   of  a  single
exponential at low $p_T$ and a power law at higher values.}
\label{fig:Fig.3}
\end{figure}
\clearpage

\begin{figure}
\vbox{\hbox to\hsize{\hfil
\epsfxsize=6.1truein\epsffile[0 0 561 751]{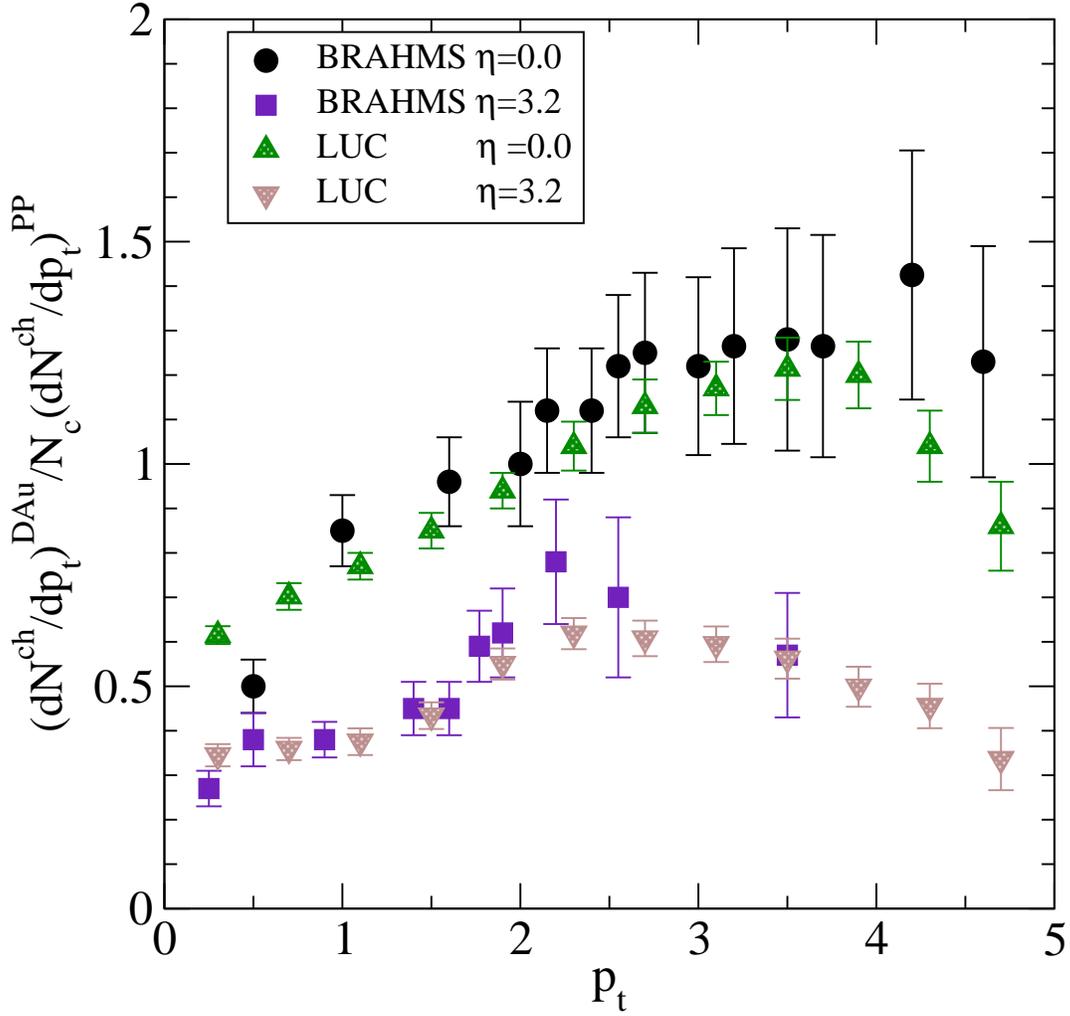}
\hfil}}
\caption[]{Minimum Bias R{DAu/PP}  for $\eta$=(0,3.2): The BRAHMS
results   are  compared   to   the  collision   number-normalized
calculations.  The latter are obtained using the results in Fig.3
with  $N_{coll}$=7.0, compared  to the  BRAHMS choice  $  7.2 \pm
0.3$. The presence  of a  Cronin effect is  clear, with  however a
flatter $p_T$ dependence obtaining for the larger $\eta$.}
\label{fig:Fig.4}
\end{figure}
\clearpage

\begin{figure}
\vbox{\hbox to\hsize{\hfil
\epsfxsize=6.1truein\epsffile[0 0 561 751]{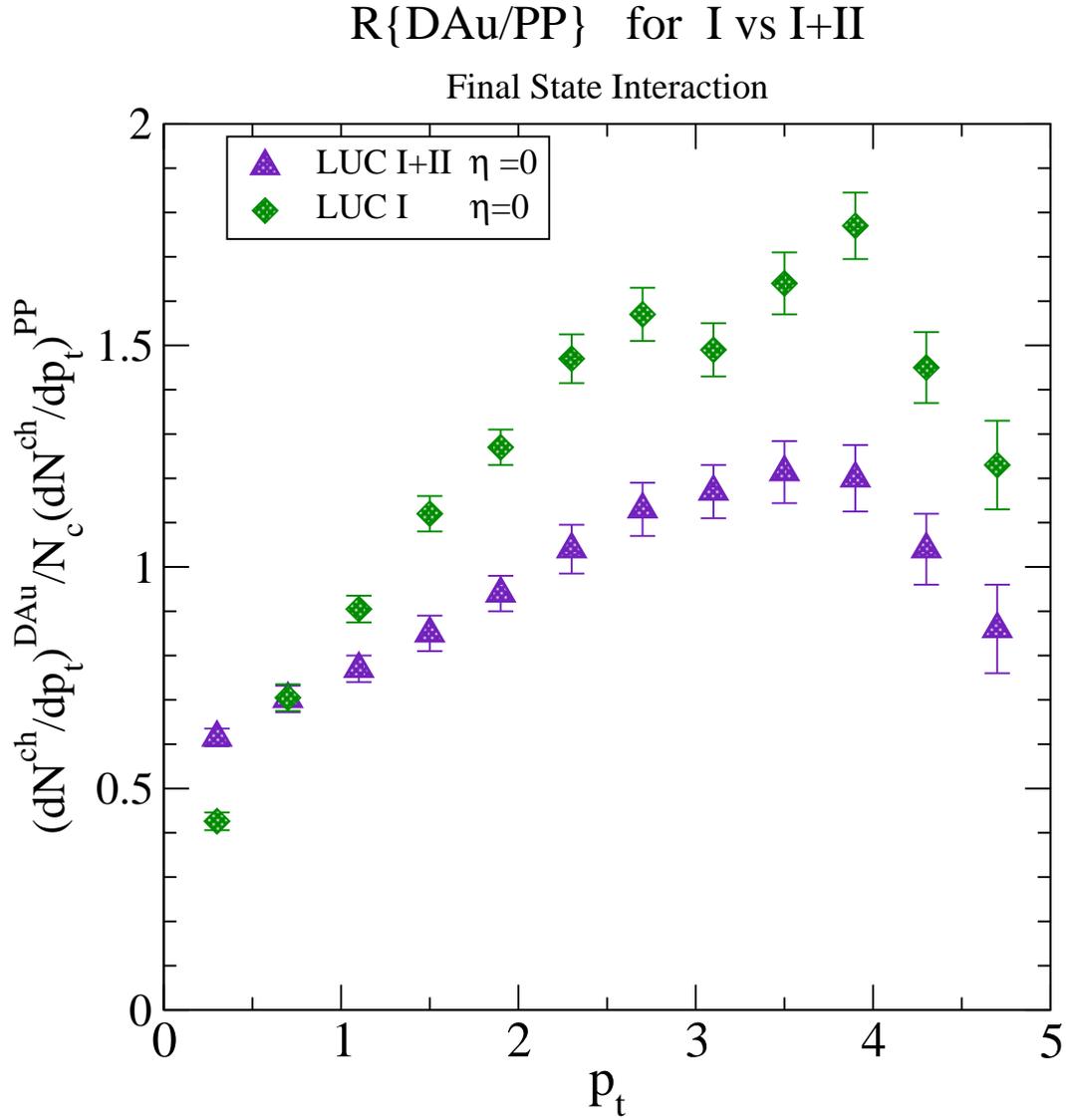}
\hfil}}
\caption[]{Effect of the Final State Hadronic Cascade. The Cronin
enhancement of $d^2N/d\eta dp_\perp$  for $\eta$=0 is markedly reduced
by the second stage of LUCIFER.  For larger $p_\perp$ one might refer
to  this  phenomenon  as  final state  `jet'  suppression.  The
diminuition is muted for  larger $\eta$. For more massive ion-ion
collisions one can expect a considerably greater reduction.}
\label{fig:Fig.5}
\end{figure}
\clearpage

\begin{figure}
\vbox{\hbox to\hsize{\hfil
\epsfxsize=6.1truein\epsffile[0 0 561 751]{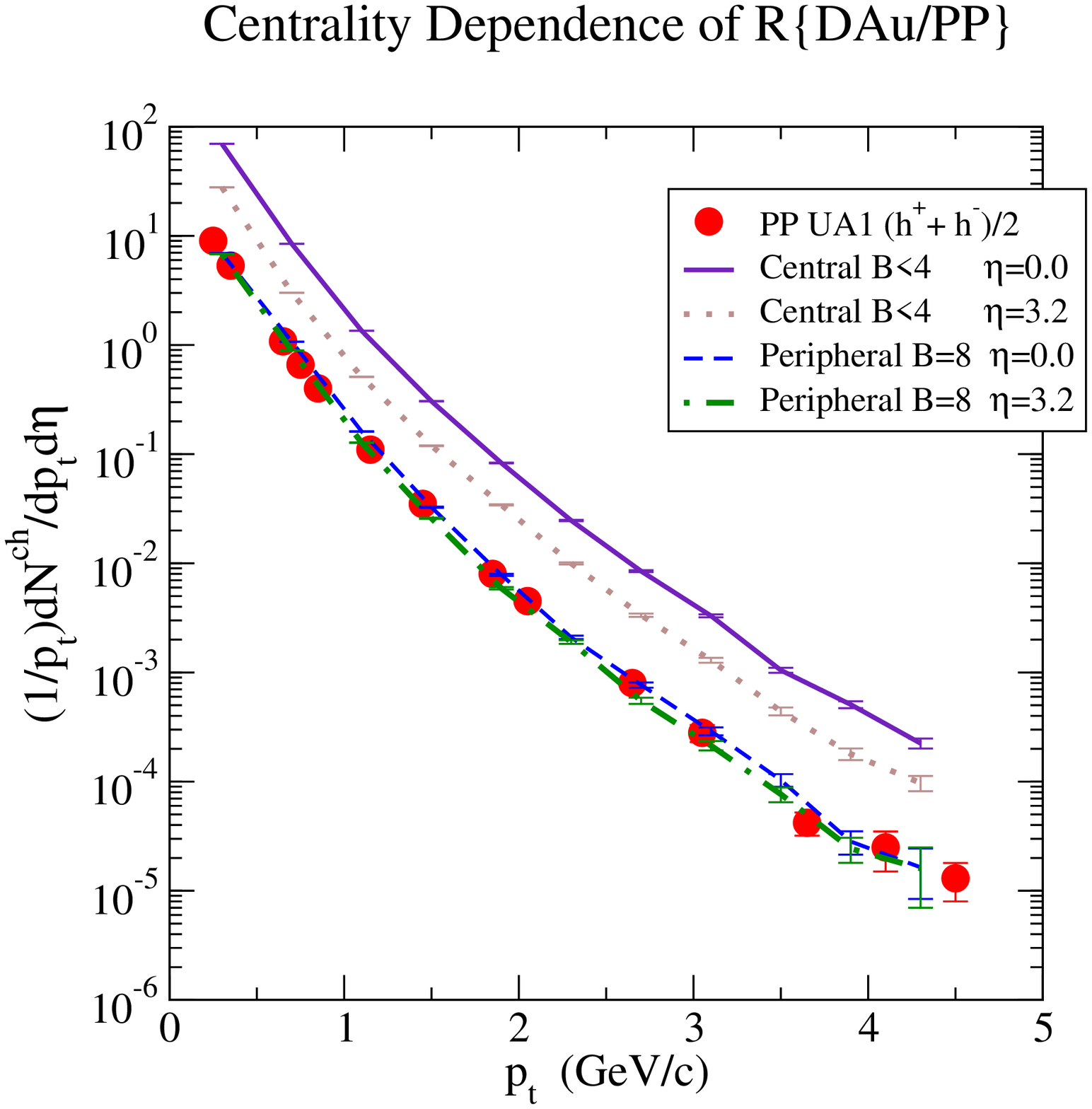}
\hfil}}
\caption[]{Centrality Dependence  of R{DAu/PP}.  The  two sets of
$\eta$=(0,3.2) transverse momentum  distributions for one central
$b \le 4.0$ and one  more peripheral impact parameter $b$=8.0 are
displayed  against   the  UA1  PP  data.   Evidently  the  $\eta$
dependence  is  strong  for  the  central  choice  and  virtually
vanishing   for  the  peripheral   collision,  with   the  latter
distributions closely matching UA1.}
\label{fig:Fig.6}
\end{figure}
\clearpage


\begin{thebibliography}{99}

\bibitem{ua5}

G.~Ekspong for the UA5 Collaboration, Nucl.~Phys.A{\bf 461}, 145c
(1987); G.~J.~~Alner     for    the     UA5    Collaboration,
Nucl.~Phys.B{\bf 291}, 445 (1987).

\bibitem{phobos1}

B.~B.~Back    {\it   et    al.},   the    PHOBOS   Collaboration,
Phys.~Rev.~Lett.,{\bf  91}  072302-1(2003):  B.~B.~Back  {\it  et
al.}, the PHOBOS Collaboration, nucl-ex/0311009, Nov.  2003;

\bibitem{brahms1}

I.~Arsene    {\it   et    al.},    the   BRAHMS    Collaboration,
nucl-ex/0307003;   I.~Arsene    {\it   et   al.},    the   BRAHMS
Collaboration,  nucl-ex/0403050;  I.~Arsene  {\it  et  al.},  the
BRAHMS  Collaboration, nucl-ex/0307003;  I.~Arsene {\it  et al.},
the BRAHMS Collaboration, nucl-ex/04030050.

\bibitem{cronin}

J.~W.Cronin {\it et al.}, Phys.~Rev.D{\bf 91}, 3105 (1979).

\bibitem{lucifer1}

D.~E.~Kahana  and  S.~H.~Kahana,  {\it Proceedings,  RHIC  Summer
Study'96},175-192,  BNL,   July  8-19,  1996;   D.~E.~Kahana  and
S.~H.~Kahana, Phys.~Rev. C{\bf  58},3574 (1998); Phys.~Rev. C{\bf
59},1651 (1999).

\bibitem{lucifer2}

D.~E.~Kahana, S.~H.~Kahana, Phys.~Rev.~C{\bf 63}, 031901(2001).

\bibitem{luc3}

Proc. International Conference on  the Physics of the Quark-Gluon
Plasma, Ecole Polytechnique,  Palaiseau, France, Sept. 4-7, 2001;
D.~E.~Kahana, S.~H.~Kahana, nucl-th/0208063.

\bibitem{rqmd}

H.~Stoecker  and  W.~Greiner, Phys.~Rep.  {\bf  137}, 277  (1986;
R.~Matiello,      A.~Jahns,     H.~Sorge      and     W.~Greiner,
Phys.~Rev.~Lettt.,{\bf 74}, 2180 (1995).

\bibitem{rqmd2}

H.~Sorge, Phys.~Rev. C{\bf 52}, 3291 (1995).

\bibitem{bass1}

S.~A.~Bass {\it et al.}, Nucl.~Phys. A{\bf 661}, 205 (1999).

\bibitem{frithjof}

B.~Andersson,   G.~Gustafson,   G.~Ingleman,  and   T.~Sjostrand,
Phys.~Rep  {\bf 97}, 31  (1983); B.~Andersson,  G.~Gustafson, and
B.~Nilsson-Almqvist,  Nucl.~Phys B{\bf 281}, 289 (1987).

\bibitem{capella}

A.~Capella and  J.~Tran Van,  Phys.~ Lett.~B{\bf 93},  146 (1980)
and Nucl.~Phys.~A{\bf 461}, 501c (1987); A.~Capella {\it et al.},
nucl-th/0405067 and hep-ph/0403081.

\bibitem{werner}

K.~Werner,  Z.~Phys. C{\bf  42},   85  (1989);  K.~  Werner,  J.~
Aichelin,  Phys.~Rev.~Lett.  {\bf   76}  (1996)  1027-1030;  H.~J.~
Drescher,  M.~Hladik,  S.~Ostapchenko,  K.~Werner, Proc.  of  the
``Workshop   on   Nuclear   Matter   in  Different   Phases   and
Transitions'', Les  Houches, France, March  31 - April  10, 1998;
K.~Werner  Invited lecture,  given at  the  Pan-American Advanced
Study Institute  "New States of Matter  in Hadronic Interactions"
Campos de Jordao, Brazil, January 7-18,2002, hep-ph/0206111.

\bibitem{ko}

B.~Zhang,  C.~M.~Ko, B-A,~Li, Z.~Lin,  nucl-th/9904075; Zi-wei~Li
and C.~m.~Ko,Phys.Rev.  C{\it 68}, 054904  (2003); Zi-wei~Li {\it
et al.}  Nucl.Phys. A {\bf 698}, 375-378 (2002).

\bibitem{ranft}

J.~Ranft  and  S.~Ritter,  Z.~Phys.~C{\bf  27}, 413  (1985);
J.~Ranft   Nucl.~Phys.A{\bf 498},  111c (1989). 


\bibitem{boal} D.~Boal, {\it Proceedings of the RHIC Workshop I},
(1985) and Phys.~Rev.C{\bf 33}, 2206 (1986).

\bibitem{eskola}

K.~J.~Eskola,  K.~ Kajantie  and J.~Lindfors,  Nucl.~Phys.  B{\bf
323}, 37 (1989).

\bibitem{wang}

X.~-N.~Wang and  M.~Gyulassy, Phys.~Rev. D{\bf  44}, 3501 (1991);
X.~-N.~Wang,  {nucl-th/000814}  and nucl-th/0405029;  X.~-N.~Wang
hep-ph/0405125.

\bibitem{geiger}

K.~Geiger and B.~Mueller Nucl.~Phys.  B{\bf 369}, 600 (1992); K.~
Geiger   Phys.~Rev.    D{\bf   46},   4965   and   4986   (1992);
K.~Geiger,{\it  Proceedings  of  Quark Matter'83},  Nucl.~Phys.~A
{\bf 418},  257c (1984);  K.~Geiger, Phys.~Rev.~D {\bf  51}, 2345
(1995).

\bibitem{bass2}

S.~A.~Bass,  B.~Mueller  and D.~K.~Srivastava,  Phys.~Rev.~Lett.B
{\it 551},277 (2003); S.~A. Bass {\it et al.}, Nucl.~Phys. A{\bf
661}, 205 (1999).

\bibitem{greiner}

K.~Gallmeister,  C.~Greiner  and  Z.~Xu, Phys.~Rev.~C{\it  67},
044905 (2003).

\bibitem{cassing}

W.~Cassing,  K.~Gallmeister, C.~Greiner Nucl.Phys.  A735, 277-299
(20004); J.~Geiss,  C.~Greiner,  E.~Bratkovskaya,  and  U.~Mosel,
Phys.~Lett. B{\bf 447}, 31 (1999).

\bibitem{goulianos}

K.~Goulianos, {\it Phys.~Rep.}{\bf 101}, 169 (1983).

\bibitem{ua1}

C.~Albajar {\it et al.}, the UA1 Collaboration, Nucl.~Phys. B{\bf
335}, 261-287 (1990). 

\bibitem{fermilab}

Y.~Eisenberg {\it  et al.},  Nucl.~Phys. A{\bf 461},  145c (1987)
G.~J.~Alner {\it  et al.}, Nucl.~Phys.  B {\bf 291},  445 (1987);
F.~Abe et al., {Phys.~Rev.}{\bf D41} 2330, (1990).

\bibitem{phobos3}

B.~B.~Back  {\it et  al.}, the  PHOBOS  Collaboration, Phys.~Rev.
~Lett. {\bf 88},22302 (2002).


\bibitem{boris}
 
B.~Z.~Kopeliovich  {\it et al.},  hep-/030120, Nucl.~Phys.  A (in
press).

\bibitem{mueller}

A.~H.~Mueller, Eur.~Phys.~J. A{\bf 1}, 19 (1998).


\bibitem{zahed}
 
E.~V.~Shuryak and I.~Zahed, hep-ph/0307267; and hep-th/0308073.

\bibitem{lattice}

S.~Datta,F.~Karsch,       P.~Petreczky      and      I.~Wetzorke,
hep-lat/0208012; hep-lat/0403017; hep-lat/0309012.

\bibitem{molnar}

D.~Molnar  and   M.~Gyulassy,  nucl-th/0102031;  nucl-th/0104018;
D.~Molnar, Talk  presented at workshop  on "Creation and  Flow of
Baryons in Hadronic and  Nuclear Collisions", ECT, Trento, Italy,
May 3-7, 2004 (unpublished).

\bibitem{flow}

S.~S.~Adler   {\it   et  al.}    ,   the  PHENIX   Collaboration,
Phys.~Rev.~Lett. {\bf 91}, 182301  (2003); C.~Adler {\it et al.},
the STAR Collaboration, Phys.~Rev.~Lett. {\bf 87}, 182301 (2001),
S.~Manly,  the  PHOBOS  Collaboration,  Proceedings of  the  20th
Winter  workshop on Nuclear  Dynamics, Trelawney  Beach, Jamaica,
March15-20, 2003.


\bibitem{gottfried}

K.~Gottfried,  Phys.~Rev.~Lett.    {\bf  32},  957   (1974);  and
Acta.~Phys.~Pol B{\bf 3}, 769 (1972).



\bibitem{saturation}

L.~V.Gribov, E.~M.Levin and M.~G.~Ryskin, Phys.~Rep. {\bf 100}, 1
(1983); Nucl.~Phys.  B{\bf 188},  555  (1981); A.~H.~Mueller  and
J.~Qiu,  Nucl.~Phys.   B{\bf  268},427  (1986);   E.~M.Levin  and
M.~G.~Ryskin, Phys.~Rep. {\bf 189}, 267 (1990).

\bibitem{cgc1}

L.~McLerran  and  R.~Venugopalan,   Phys.~Rev.  D{\bf  49},  2223
(1994); Phys.~Rev.   D{\bf  59},   094002   (1999);  D.~Kharzeev,
E.~Levin and L.~McLerran, Phys.~Lett. B{\bf 561}, 93 (2003).

\bibitem{cgc2}

D.~Kharzeev,E.~M.Levin  and  L.~Mclerran  hep-ph/0210332  (2002);
L.~Mclerran  hep-ph/0402137 (2004);  D.~Kharzeev,  E.~M.Levin and
L.~Mclerran, hep-ph/0403271 (2004).

\bibitem{wang2}

M.~Gyulassy and X.~N.~Wang, Comp.~Phys.~Comm. {\bf83}, 307
(1994), 

\bibitem{ampt}

Zei-wei Lin and C.~M.~Ko, nucl-th/0301025.



\end{thebibliography}
\end{document}